\documentclass[useAMS,usenatbib]{mn2e}

\usepackage{color}
\usepackage{graphics}
\usepackage{psfig,subfigure}
 
\title[Why are accreting T Tauri stars observed to be less luminous in X-rays than non-accretors?]
      {Why are accreting T Tauri stars observed to be less luminous in X-rays than non-accretors?}
\author[S. G. Gregory, K. Wood and M. Jardine]
{S. G. Gregory$^{1}$\thanks{E-mail: sg64@st-andrews.ac.uk}, K. Wood$^{1}$ and M. Jardine$^{1}$\\
$^{1}$SUPA, School of Physics and Astronomy, University of St Andrews, 
North Haugh, St Andrews, Fife, KY16 9SS, UK}

\begin{document}

\date{}

\pagerange{\pageref{firstpage}--\pageref{lastpage}} \pubyear{2007}

\maketitle

\label{firstpage}

\begin{abstract}
Accreting T Tauri stars are observed to be less luminous in X-rays than non-accretors, an effect that has been 
detected in various star forming regions.  To explain this we have combined, 
for the first time, a radiative transfer code with an accretion model that considers magnetic fields extrapolated 
from surface magnetograms obtained from Zeeman-Doppler imaging. Such fields consist of compact magnetic regions 
close to the stellar surface, with extended field lines interacting with the disc. We study the propagation of 
coronal X-rays through the magnetosphere and demonstrate that they are strongly absorbed by the dense gas in 
accretion columns. The reduction in the observed X-ray emission depends on the field geometry, which may explain why 
accreting T Tauri stars show a larger scatter in their observed X-ray luminosity compared to non-accreting stars.
\end{abstract}



\section{Introduction}
Accreting T Tauri stars are observed to be less luminous in X-rays than non-accretors (\citealt{ste01}; 
\citealt{fla03a,fla03c,fla06}; \citealt{sta04}; \citealt{pre05}; \citealt{tel07a}).  Accreting stars appear to be a 
factor of $\sim 2$ less luminous, and show a larger variation in their X-ray activity compared to non-accreting 
stars \citep{pre05}. However, it is only in recent years that this result has become clear, with previous studies 
showing conflicting results (e.g. \citealt{fei03} and \citealt{fla03b}).  The apparent discrepancy arose from
whether stars were classified as accreting based on the detection of excess IR emission (a disc indicator) or
the detection of accretion related emission lines.  However, with careful re-analysis of archival data 
\citep{fla03a} and recent large X-ray surveys like the Chandra Orion Ultradeep Project (COUP; \citealt{get05}) and the 
XMM-Newton Extended Survey of the Taurus Molecular Cloud (XEST; \citealt{gue07a}) the result is now clear, namely 
that accreting T Tauri stars are observed to be, on average, less luminous in X-rays than non-accreting stars.  Although the 
difference is small it has been found consistently in various star forming regions, namely Taurus-Aurigae (\citealt{ste01}; 
\citealt{tel07a}), the ONC (\citealt{fla03c}; \citealt{sta04}; \citealt{pre05}), NGC 2264 (\citealt{fla03a,fla06})
and Chamaeleon I (\citealt{fla03a}).  

It should be noted, however, that such observations from CCD detectors are not very sensitive to X-rays that are 
produced in accretion shocks.  High resolution X-ray spectroscopic measurements have indicated emission from
cool and high density plasma, most likely associated with accretion hot spots, in several (but not all) accreting stars (e.g. 
\citealt{tel07b}; \citealt{hans07}).  In this letter we only consider coronal X-ray emission such as is 
detected by CCD measurements.   

It is not yet understood why accreting stars are under luminous in X-rays, although a few ideas have been put 
forward.  It may be related to higher extinction due to X-ray 
absorption by circumstellar discs, however the COUP results do not support this suggestion \citep{pre05}.  It may 
be related to magnetic braking, whereby the interaction between the magnetic field of an accreting 
star with its disc slows the stellar rotation rate leading to a weaker dynamo action and therefore less X-ray 
emission; although the lack of any rotation-activity relation for T Tauri stars has ruled out this idea (\citealt{fla03c}; 
\citealt{pre05}; \citealt{bri07}).  A third 
suggestion is that accretion may alter the stellar structure affecting the magnetic field generation process and therefore 
X-ray emission \citep{pre05}.  However, the most plausible suggestion 
is the attenuation of coronal X-rays by the dense gas in accretion columns (\citealt{fla03c}; \citealt{sta04}; 
\citealt{pre05}; \citealt{gue07b}).  X-rays from the underlying corona may not be able to 
heat the material within accretion columns to a high enough temperature to emit in X-rays.  Field lines which have 
been mass-loaded with dense disc material may obscure the line-of-sight to the star at some rotation 
phases, reducing the observed X-ray emission.  In this letter we demonstrate this in a quantitative way
by developing an accretion flow model and simulating the propagation of coronal X-rays 
through the stellar magnetosphere.  


\section{Realistic Magnetic Fields}
\begin{figure*}
        \def\subfigtopskip{4pt}
        \def\subfigbottomskip{4pt}
        \def\subfigcapskip{2pt}
        \centering
        \begin{tabular}{c}
        \subfigure{\psfig{figure=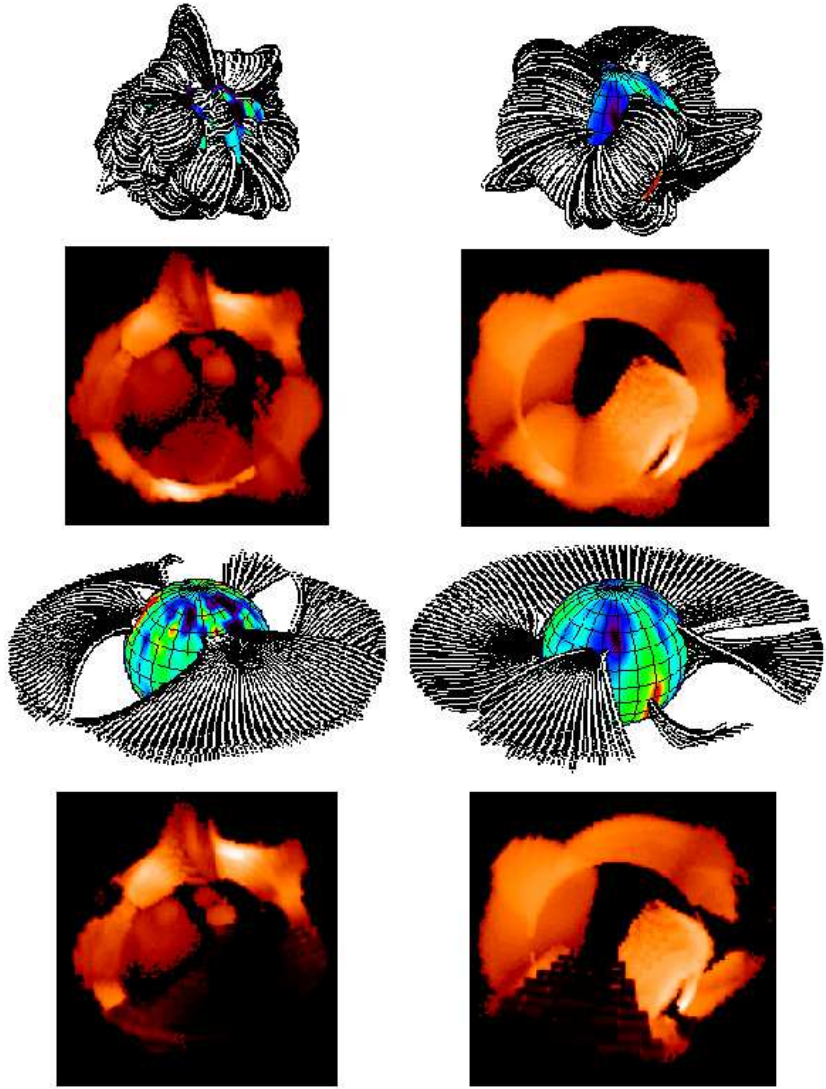,width=110mm}} 
        \end{tabular}
        \caption{The model coronal (first row) T Tauri magnetic fields extrapolated from 
                 the AB Dor (left-hand column) and LQ Hya (right-hand column) surface magnetograms, with the 
                 corresponding X-ray corona (second row) assuming a stellar inclination of $60\degr$.    
                 Also shown is the structure of the accreting field (third row) and the X-ray emission images assuming that
                 accretion is taking place (fourth row) - notice the occulted X-ray bright regions.   
                 For some lines-of-sight the X-ray bright regions are lightly obscured by the accretion columns, reducing the 
                 observed X-ray emission.  For other lines-of-sight the coronal X-rays cannot penetrate the dense accreting gas.  
                 The average reduction in the observed X-ray EM across an entire rotation cycle is factor of 1.4 (2.0) for the 
                 AB Dor-like (LQ Hya-like) field.  The images are not to scale.}
        \label{coronae}
\end{figure*}

In order to model the coronae of T Tauri stars we need to assume something about the form of the magnetic field.
Observations suggest it is compact and inhomogeneous and may vary not only with time on each star, but also
from one star to the next. To capture this behaviour, we use as examples the field structures of two different
main sequence stars, LQ Hya and AB Dor determined from Zeeman-Doppler imaging \citep{don03}.  
Although we cannot be certain whether or not the magnetic field structures extrapolated from 
surface magnetograms of young main sequence stars do represent the magnetically confined coronae of T Tauri 
stars, they do satisfy the currently available observational constraints.  In future it will
be possible to use real T Tauri magnetograms derived from Zeeman-Doppler images obtained using
the ESPaDOnS instrument at the Canada-France-Hawaii telescope \citep{don07}.  However, in the meantime, 
the example field geometries used in this letter (see Fig. \ref{coronae}) capture the essential features
of T Tauri coronae.  They reproduce X-ray emission measures (EMs) and coronal densities which 
are typical of T Tauri stars \citep{jar06}.  The surface field structures are complex, consistent with polarisation
measurements \citep{val04} and X-ray emitting plasma is confined within unevenly distributed magnetic structures 
close to the stellar surface, giving rise to significant rotational modulation of X-ray emission \citep{gre06b}.  

We extrapolate from surface magnetograms by assuming that the magnetic field $\bmath{B}$ is potential such that 
$\nabla \times \bmath{B} = 0$. This condition is satisfied 
by writing the field in 
terms of a scalar flux function $\Psi$, such that $\bmath{B}=-\nabla \Psi$.  Thus, in order to ensure that the 
field is divergence-free ($\nabla.\bmath{B}=0$), $\Psi$ must satisfy Laplace's equation, 
$\nabla^2\Psi=0$; the solution of which is a linear combination of spherical harmonics,
\begin{equation}
\Psi= \sum_{l=1}^{N} \sum_{m=-l}^{l} \left
[a_{lm}r^l+b_{lm}r^{-(l+1)} \right ] P_{lm}(\theta) {\rm e}^{{\rm i}m\phi},
\end{equation}
where $P_{lm}$ denote the associated Legendre functions.  The coefficients $a_{lm}$ and $b_{lm}$ 
are determined from the radial field at the stellar surface obtained from Zeeman-Doppler maps and 
also by assuming that at some height $R_s$ above the surface (known as the source surface) the 
field becomes radial and hence $B_{\theta}(R_s)=0$, emulating the effect of the corona blowing 
open field lines to form a stellar wind \citep{alt69}.  In order to extrapolate the field we used 
a modified version of a code originally developed by \citet*{van98}.


\subsection{The coronal field}
For a given surface magnetogram we calculate the extent of the closed corona for a specified
set of stellar parameters.  As this process has been described in detail by \citet{jar06} and 
\citet{gre06a,gre06b} we provide only a brief outline here.  We assume that the
corona is isothermal and that plasma along field line loops is in hydrostatic equilibrium.  
The pressure is calculated along the path of field lines loops and is set to zero for open field 
lines and for field lines where, at some point along the loop, the gas pressure exceeds the 
magnetic pressure.  The pressure along a field line scales with the pressure at its 
foot point, and we assume that this scales with the magnetic pressure ($p_0=KB_0^2$).      
This technique has been used successfully to calculate mean coronal densities and X-ray EMs for 
the Sun and other main sequence stars \citep{jar02} as well as T Tauri stars \citep{jar06}.   
The extent of the corona depends both on the value of $K$ and also on $B_0$ which is determined 
directly from surface magnetograms and varies across the stellar
surface.  \citet{jar06} determined the best value of $K$ for a given surface magnetogram which results in the 
best fit to the X-ray EMs of stars from the COUP database.  We have adopted the same values 
in this letter.  We note that we make a conservative estimate of the location of the source surface by calculating 
the largest radial distance at which a dipole field line would remain closed, with the same average field strength.   
The AB Dor-like coronal field has an X-ray EM\footnote{The X-ray EM is given by $EM =\int n^2dV$ where $n$ and $V$
are the coronal density and volume.  The EM-weighted density is $\bar{n}=\int n^3dV/\int n^2dV$.} of 
$\log{EM}=53.73\,{\rm cm}^{-3}$ (without considering accretion) and a mean EM-weighted coronal density of 
$\log{\bar{n}}=10.57\,{\rm cm}^{-3}$, consistent with estimates from the modelling 
of individual flares \citep{fav05}.  The LQ Hya-like field has a more extended corona and consequently a lower 
coronal density and EM, $\log{EM}=52.61\,{\rm cm}^{-3}, \log{\bar{n}}=9.79\,{\rm cm}^{-3}$.

\subsection{The accreting field}
We consider a thin disc and assume that the disc normal is aligned with the stellar
rotation axis.  We consider a steady state flow model and assume that the structure
of the magnetic field remains undistorted by the in-falling material and that the magnetosphere
rotates as a solid body.  \citet{jar06} and \citet{gre06a} demonstrated that for most stars the closed field region
is confined close to the stellar surface.  T Tauri stars have hot coronae, with temperatures
in excess of several tens of mega-Kelvin [e.g. see the COUP database, \citet{get05}].  Such
high temperatures lead to closed field lines being opened by the gas pressure, limiting 
the closed field region to close to the stellar surface.  In 
such cases the inner disc may sit in a reservoir of open field \citep{gre06a}.  This can be seen
in Fig. \ref{coronae} where the field lines that are carrying accreting gas are open.          
The question of where the disc is truncated remains a major problem for accretion models.   It is still 
unknown if the disc is truncated in the vicinity of the corotation radius, the assumption of traditional 
accretion models (e.g. \citealt{kon91}), or whether it extends closer to the stellar surface
(e.g. \citealt{mat05}).  For this letter we consider that accretion occurs over a range of radii within the corotation 
radius.  This is equivalent to the approach taken previously 
by \citet*{har94}, \citet*{muz01}, \citet*{sym05} and \citet{aze06} who have 
demonstrated that such an assumption reproduces 
observed spectral line profiles and variability.  It should also be noted that        
the accreting field geometries which we consider here are only snap-shots in time, and in reality will
evolve due to the interaction with the disc.  The accretion filling factors are $0.9\%$ for the 
AB Dor-like field and $1.2\%$ for the LQ Hya-like field, smaller 
than would be expected for accretion to a dipole, but consistent with observationally inferred values 
(e.g. \citealt{val04}).  

We assume that material is supplied by the disc and accretes onto the star at a constant rate.
For a dipolar magnetic field accretion occurs into two rings in opposite hemispheres centred on
the poles.  In this case, half of the mass supplied by the disc accretes into each
hemisphere.  For complex magnetic fields accretion occurs into discrete hotspots 
distributed in latitude and longitude \citep{gre06a}.  It is therefore not clear how 
much of the available mass from the disc accretes into each hot spot.  
We use a spherical grid and assume that each grid
cell within the disc which is accreting supplies a mass accretion rate that is proportional
to its surface area.  For example, if an accreting grid cell has a surface area that is $1\%$ of the 
total area of all accreting grid cells, then this grid cell is assumed to carry $1\%$ of the 
total mass that is supplied by the disc.  Therefore, as an example, if grid cells
which constitute half of the total area of all accreting cells in the disc carry material into a 
single hotspot, then half of the mass accretion rate is carried from the disc to this hot spot.  In 
this way the accretion rate into each hot spot is different and depends on the structure of
the magnetic field connecting the star to the disc.


\subsection{Accretion flow model}
\begin{figure}
  \centering
  \psfig{file=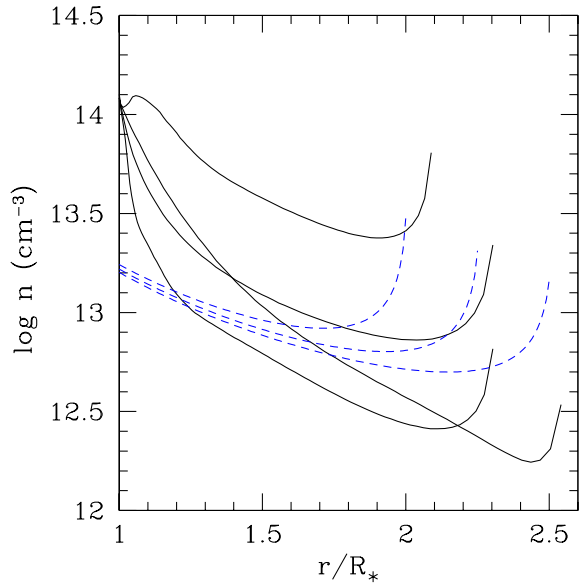,width=80mm}
  \caption{Some example density profiles (solid lines) for accretion along a small selection of the
           complex field lines shown in Fig. \ref{coronae} (lower left panel) assuming a mass accretion
           rate of $10^{-7}\,{\rm M}_{\odot}\,{\rm yr}^{-1}$.  Also shown for comparison are the density profiles for
           accretion along dipolar field lines (dashed lines). $r$ is the spherical radius.}
  \label{flows}
\end{figure}

We consider a star of mass $0.5\,{\rm M}_{\odot}$, radius $2\,{\rm R}_{\odot}$,
rotation period $6\,{\rm d}$, a coronal temperature of $20\,{\rm MK}$ and assume
that the disc supplies a mass accretion rate of $10^{-7}\,{\rm M}_{\odot}\,{\rm yr}^{-1}$.
In order to model the propagation of coronal X-rays through the magnetosphere 
we first need to determine the density of gas within accretion columns.  
Provided that the flow of accreting material is along the path of the 
field ($\bmath{v}$ is parallel to $\bmath{B}$) then
\begin{equation}
\frac{\rho v}{B} = const.
\label{mass}
\end{equation} 
where $\rho$ is the mass density (see e.g. \citealt{mes68}).  If an individual accreting 
field line carries material onto an area of the stellar surface 
of $A_{\ast}$ with velocity $v_{\ast}$ and density $\rho_{\ast}$, then the mass accretion rate into
$A_{\ast}$ is,
\begin{equation}
\dot{M} = \rho_{\ast} v_{\ast} A_{\ast}.
\label{mdot}
\end{equation}
It then follows from (\ref{mass}) and (\ref{mdot}) that the density profile along the path 
of the field line may be written as,
\begin{equation}
\rho(r) = \frac{B(r)}{B(R_{\ast})}\frac{\dot{M}}{A_{\ast}v(r)}.
\end{equation} 
The density profiles therefore do not depend on the absolute field strength, but instead on
how the field strength varies with height above the star.  The density profiles are 
typically steeper than those derived for accretion flows along dipolar field lines since the 
strength of a higher order field drops faster with height above the star (see Fig. \ref{flows}).  
Gas is assumed to free-fall under gravity along the path of the field after leaving the disc with a 
small initial velocity of $10\,{\rm kms}^{-1}$ \citep{muz01}.  Accretion occurs from a range of 
radii as discussed in \S2.2.  For the AB Dor-like field accretion occurs from $\sim 2.0-2.5R_{\ast}$, 
whilst for the LQ Hya-like field, which is more extended, from $\sim 2.4-2.9R_{\ast}$.

In order to estimate the number density of gas within the accretion columns we 
require the temperature distribution along the path of the field lines.  The temperature
variation within the accretion columns of T Tauri stars is uncertain.  The most
comprehensive model is that of \citet{mar96} who considers both heating
and cooling processes.  Here we elect to take the simplest approach and assume an 
isothermal accretion flow temperature of $7000\,{\rm K}$ - a reasonable estimate for the 
temperature of accreting material (see e.g. \citealt{har94}; \citealt{mar96}).  We obtain the 
number density profiles from,
\begin{equation}
n(r) = \rho(r)/\mu m_H,
\end{equation}    
where $m_H$ is the mass of a hydrogen atom and $\mu$ the dimensionless atomic weight.  Fig. \ref{flows}
shows the variation of the number density along the paths of a selection of accreting field lines, with 
those obtained for dipolar field lines shown for comparison.  For our assumed accretion rate of 
$10^{-7}\,{\rm M}_{\odot}\,{\rm yr}^{-1}$ the flow densities range from $\log {n} \approx 12-14\,{\rm cm}^{-3}$, whilst for a 
lower accretion rate of $10^{-8}\,{\rm M}_{\odot}\,{\rm yr}^{-1}$ the range is $\log {n} \approx 11-13\,{\rm cm}^{-3}$.  


\section{Simulated X-Ray Variability}
We extend our previous work on optically thin X-ray emission from the
coronae of young stars (e.g. \citealt{jar02,jar06}; \citealt{gre06b}) to include
the effects of absorption by the dense accretion columns.  In this model we assume
the X-ray emission from the 20~MK coronae is optically thin, but that
the X-rays may be subsequently absorbed in the cool (7000~K) and hence
optically thick accretion columns.  For the X-ray absorptive opacity we
adopt a value of $\sigma = 10^{-22}\,{\rm cm}^2\,{\rm H}^{-1}$,
typical of neutral gas at temperatures below $10^4$~K at X-ray energies
of a few keV (e.g. \citealt{kro84}).  At these energies the
opacity of hot gas (above $10^7$~K) is several orders of magnitude
lower (e.g. \citealt{kro84}, Fig. 1) justifying our
assumption that the 20~MK coronal X-ray emission is optically thin.

As in our previous work, for the radiation transfer we use Monte Carlo
techniques and discretise the emissivity and density onto a
spherical polar grid  (e.g. \citealt{jar02};
\citealt{whi03}).  In our simulations we adopt a stellar
inclination of $i=60\degr$.  For inclinations above $i\ga 75\degr$
the star and hence coronal emission, will be blocked by the vertically
extended (i.e. flared) discs typical of T Tauri stars (e.g. \citealt{dal98}; \citealt{woo02}).  
Because we are focusing on inclinations where the star is not blocked by the disc, our simulations
only require our emissivity and density grid to include the accretion
columns and therefore our grid extends to the inner edge of the disc.
In the Monte Carlo X-ray radiation transfer simulations we assume the
scattering opacity is negligible, so our results in Fig. \ref{coronae} show the
effects of attenuation of the coronal emission by the accretion
columns.  The second row in Fig. \ref{coronae} shows the X-ray images in the 
absence of attenuation (i.e. X-ray opacity in the accretion columns is set to zero) whilst the fourth row 
shows the same X-ray emission models, but with our adopted value
for the soft X-ray opacity in the accretion columns.

For the AB Dor-like field structure the observed X-ray EM is reduced by a factor of 1.4
when accretion flows are considered, whereas for the LQ Hya-like field the reduction is
by a factor of 2.0 (where the reduction factor is the average for an entire rotation cycle).  
For the AB Dor-like field there are large accretion curtains which cross the observers 
line-of-sight to the star as it rotates (see Fig. \ref{coronae}).  For the LQ Hya-like field accretion is 
predominantly along field lines which carry material into low latitude hot spots, however, one of the brightest
X-ray emitting regions is obscured by an accretion column which attenuates the coronal X-rays and produces
a large reduction in the observed X-ray emission.  This immediately suggests that the geometry of the 
accreting field is a contributory factor in causing the large scatter seen in the X-ray luminosities of 
accreting stars.  


\section{Conclusions}
We have demonstrated that the suppression of X-ray emission in accreting stars apparent from CCD observations
can, at least in part, be explained by the attenuation of coronal X-rays by the dense material in accretion
columns. The reduction in the observed X-ray emission depends on the structure of the accreting field.  
For stars where accretion columns rotate across the line-of-sight, X-rays from the underlying corona 
are strongly absorbed by the accreting gas which reduces the observed X-ray emission.  This however does not rule 
out the fact that other mechanisms may also be responsible for reducing the X-ray emission in accreting stars.  
\citet{jar06} have demonstrated that some stars (typically those of lower mass)
have their outer coronae stripped away via the interaction with a disc.  This also reduces the observed X-ray
emission and this effect, combined with the radiative transfer calculations presented in this letter,
is likely to lead to a larger reduction in the observed X-ray emission.  This would reduce the number of field lines
which could be filled with coronal gas, such as is also suggested by \citet{pre05} and \citet{tel07a}, with
the observed X-ray emission being further reduced due to obscuration by the accreting gas.  A subsequent paper 
will give a detailed analysis of both of these effects drawing on X-ray data from the COUP and XEST.


\section*{Acknowledgements}
The authors thank the referee for positive comments and Ad van Ballegooijen who wrote the original version of the 
potential field extrapolation code.

\bibliographystyle{mn2e}
\bibliography{flows}

\bsp

\label{lastpage}

\end{document}